\documentclass[prd,twocolumn,superscriptaddress]{revtex4-1}
\usepackage{epsfig}
\usepackage{graphicx}
\usepackage{palatino}
\usepackage[english]{babel}
\usepackage{hyphenat}
\usepackage{amsmath}
\usepackage{amssymb}
\usepackage{mathtools}
\usepackage{mathrsfs}
\usepackage{slashed}
\usepackage{epstopdf}
\usepackage{xcolor}
\usepackage{booktabs}
\definecolor{lcolor}{rgb}{0.,0.0,0.}
\definecolor{citcolor}{rgb}{0,0.,0.5}
\usepackage[breaklinks,colorlinks,urlcolor=blue,citecolor=blue,linkcolor=blue]{hyperref}
\usepackage{multirow}
\usepackage{ltablex}
\usepackage{soul}

\newcommand{\beq}{\begin{eqnarray}}
\newcommand{\eeq}{\end{eqnarray}}

\newcommand{\bem}{\begin{multline}}
\newcommand{\eem}{\end{multline}}
\newcommand{\beg}{\begin{gather}}
\newcommand{\eeg}{\end{gather}}

\newcommand{\nn}{\nonumber\\}

\newcommand{\ben}{\begin{eqnarray*}}
\newcommand{\een}{\end{eqnarray*}}

\def\cO{{\cal O}}

\newcommand{\eqn}[1]{Eq.~\eqref{#1}}

\newcommand{\secn}[1]{Section~1}
\newcommand{\appn}[1]{Appendix~1}

\long\def\comment#1{ }

\def\and{\quad\text{and}\quad}

\newcommand{\dip}{{\rm dip}}
\newcommand{\dif}{{\rm d}}
\newcommand{\rmd}{{\rm d}}

\newcommand{\rme}{{\rm e}}

\newcommand{\qhat}{\hat{q}}
\newcommand{\vect}[1]{\boldsymbol{#1}_{\perp}}

\newcommand{\kt}{\vect{k}}
\newcommand{\ktg}{\boldsymbol{k}_{g\perp}}

\newcommand{\xt}{\vect{x}}
\newcommand{\qt}{\vect{q}}
\newcommand{\zt}{\vect{z}}
\newcommand{\yt}{\vect{y}}
\newcommand{\rt}{\vect{r}}

\def\0{{\boldsymbol 0}}

\newcommand{\del}{\partial}

\def\rmd{\textrm{d}}

\newcommand{\abar}{\bar{\alpha}_s}

\begin{document}

\title{Transverse momentum broadening from NLL BFKL to all orders in pQCD }
\author{Paul Caucal} 
\email{caucal@subatech.in2p3.fr}
\affiliation{Physics Department, Brookhaven National Laboratory, Upton, NY 11973, USA}
\affiliation{SUBATECH UMR 6457 (IMT Atlantique, Universit\'e de Nantes, IN2P3/CNRS), 4 rue Alfred Kastler, 44307 Nantes, France}
\author{Yacine Mehtar-Tani} 
\email{mehtartani@bnl.gov}
\affiliation{Physics Department, Brookhaven National Laboratory, Upton, NY 11973, USA}
\affiliation{RIKEN BNL Research Center, Brookhaven National Laboratory, Upton, NY 11973, USA}

\begin{abstract}
We study, to all orders in perturbative QCD, the universal behavior of the saturation momentum $Q_s(L)$ controlling the transverse momentum distribution of a fast parton propagating through a dense QCD medium with large size $L$. Due to the double logarithmic nature of the quantum evolution of the saturation momentum, its large $L$ asymptotics is obtained by slightly departing from the double logarithmic limit of either next-to-leading log (NLL) BFKL or leading order DGLAP evolution equations. At fixed coupling, or in conformal $\mathcal{N}=4$ SYM theory, we derive the large $L$ expansion of $Q_s(L)$ up to order $\alpha_s^{3/2}$. In QCD with massless quarks, where conformal symmetry is broken by the running of the strong coupling constant, the one-loop QCD $\beta$-function fully accounts for the universal terms in the $Q_s(L)$ expansion. Therefore, the universal coefficients of this series are known exactly to all orders in $\alpha_s$.
\end{abstract} 

\maketitle

\section{Introduction}

The study of the suppression and modification of jets in heavy-ion collisions at RHIC and LHC \cite{PHENIX:2001hpc,PHENIX:2003qdj,STAR:2002ggv,STAR:2002svs,STAR:2003fka,ALICE:2012aqc,ALICE:2013dpt,CMS:2012aa,ATLAS:2014ipv}, commonly referred to as ``jet quenching" \cite{Blaizot:2015lma,Qin:2015srf,Cunqueiro:2021wls}, aims at probing the quark-gluon plasma at various scales as well as non-equilibrium dynamics of QCD. This physics will play a central role in sPHENIX program at RHIC \cite{PHENIX:2015siv} and the upcoming run 4 at the LHC \cite{Citron:2018lsq}. In that context, the phenomenon of Transverse Momentum Broadening (TMB) of jets in the QGP is of major importance as it controls many jet quenching related observables measured in heavy-ion collisions. For instance, TMB is responsible for the jet energy loss by deflecting soft medium-induced gluons at larger angles than the jet cone size $R$ \cite{Blaizot:2013hx} leading to the suppression of the jet cross-section in nucleus-nucleus collisions \cite{ATLAS:2018gwx,ALICE:2019qyj,Caucal:2019uvr,Mehtar-Tani:2021fud}.  Another important and historical signature of jet quenching is the dijet azimuthal asymmetry \cite{PHENIX:2007yjc,ATLAS:2010isq,ALICE:2015mdb,STAR:2017hhs}. This observable is believed to be sensitive to TMB of jets propagating in the quark-gluon plasma through the suppression of the back-to-back peak that signals the azimuthal de-correlation of the di-jet system \cite{Mueller:2016gko,Chen:2016vem}. Recently, the possibility of measuring TMB using jet substructure observables like the Soft Drop grooming angle has also been investigated \cite{Ringer:2019rfk,STAR:2021kjt}.

TMB is encoded in the so-called jet quenching parameter $\qhat$, which is roughly speaking the average transverse momentum squared $\kt^2$ acquired per unit of time by a fast parton propagating in a dense medium 
\beq 
\qhat\sim\frac{\rmd \langle \kt^2\rangle} {\rmd  t}\,.  \label{eq:qhat-simpledef}
\eeq
In the absence of quantum corrections, this simple brownian diffusion in transverse space picture leads to a TMB distribution which is peaked around a characteristic transverse momentum scale, the saturation scale $Q_s$ such that 
\beq \label{eq:sat-scale}
Q_s^2\sim \qhat L\,,
\eeq 
for a given system size $L$. On the other hand, at large $\kt$, the distribution becomes dominated by rare, single hard scattering with a medium quasi-particle, and therefore displays the characteristic Rutherford like power law $1/\kt^4$. The latter has received a lot of attention in the past few years, as it would signal the presence of point-like quasi-particles in the quark-gluon plasma \cite{DEramo:2012uzl,DEramo:2018eoy}. Of course, the study of this regime in the context of jets in heavy ion collisions is rather challenging due to the large background of soft particles and the interplay with inelastic higher order processes, on the one hand, and to the low cross-section associated with these rare events on the other hand. 

On the theory side, a lot of progress has been made in the description of TMB. A compact formula \`{a} la Moli\`{e}re \cite{Moliere+1948+78+97} encompassing both the multiple soft scattering and single hard scattering regimes has been derived in \cite{Barata:2020rdn}. Furthermore, the effects of transverse flow on the TMB distribution have been investigated analytically and numerically in \cite{Sadofyev:2021ohn,Barata:2022krd,Andres:2022ndd}. Also, in the small transverse momentum domain of the distribution, where non-perturbative physics dominates, recent lattice calculations opened up the possibility of achieving a comprehensive picture of TMB at all scales \cite{Moore:2019lgw,Moore:2021jwe}.

Quantum corrections to TMB have also triggered many recent studies since the seminal papers \cite{Liou:2013qya,Blaizot:2013vha} in which the authors show that radiative corrections are enhanced by a double logarithm of the system size $L$, $\qhat_{\rm NLO}\sim \qhat_{\rm LO}\times \alpha_s\ln^2L$. These quantum corrections of order $\alpha_s\propto g^2$ are of a different origin than the classical corrections of order $g$ associated with soft thermal modes in the plasma \cite{Caron-Huot:2008zna} (for a recent discussion about the interplay between quantum and classical corrections to $\qhat$, see \cite{Ghiglieri:2022gyv}). The logarithmic dependence on the system size encoded in higher order corrections is an expression of the non local nature of the quantum fluctuations that were shown to qualitatively  affect the underlying diffusive process. 

Remarkably, these quantum corrections were shown to be process-independent up to single logarithmic accuracy (at large $N_c$), since they also appear in NLO corrections to the medium-induced gluon spectrum \cite{Blaizot:2014bha,Iancu:2014kga,Arnold:2021mow,Arnold:2021pin}. This suggests a renormalization group approach in order to resum these potentially large logarithmic terms. Upon resummation to all orders, the resulting TMB distribution showcases interesting physical properties: the typical width of the distribution, parametrically given by the saturation momentum $Q_s$, grows with the system size $L$ faster than the standard diffusive exponent $L^{1/2}$, and the large-$\kt$ tail is a power law which deviates from the Rutherford behaviour by a $\sqrt{\alpha_s}$ correction \cite{Caucal:2021lgf}. This features are characteristic of anomalous (super) diffusive processes.  Moreover, TMB distribution and $Q_s$ tend to universal limits at large system sizes, such that these quantities are not anymore sensitive to the details of the initial conditions and the non-perturbative regime.

So far, the resummation of higher orders of $\qhat$ and the TMB distribution have been obtained at double logarithmic accuracy (DLA), meaning that only terms of the form $\alpha_s^n\ln^{2n}L$ are included to all orders in perturbation theory. 

In this paper, we address for the first time the effect of single logarithmic corrections on the saturation momentum $Q_s(L)$ in the asymptotic regime where $L$ is large. Relying on the mapping between the evolution equation for $\qhat$ and the equations governing the propagation of traveling wave fronts in non-linear physics \cite{dee1983propagating,van1987dynamical,Munier:2003vc,2000,2003}, we can compute the large $L$ asymptotic behaviour of the saturation momentum, even without having analytic control of the full non-linear evolution equation \cite{Caucal:2021lgf,Caucal:2022fhc}. We argue that the evolution of $Q_s$ is dominated by the double logarithmic regime of QCD, namely by soft and collinear gluon radiations. Therefore, the corrections beyond the DLA can be obtained either from a BFKL \cite{Kuraev:1977fs,Balitsky:1978ic} or a DGLAP \cite{Gribov:1972ri,Altarelli:1977zs,Dokshitzer:1977sg} approach, as the double logarithmic regime is common to these two equations. We choose to proceed using the BFKL equation at leading and next-to-leading logarithmic accuracy (the DGLAP case is presented in appendix~\ref{app:dglap}), as it is more easily justified from a physical point of view and simpler to implement in practice. 

The main results of the paper are the universal asymptotic behaviour of $Q_s$ at three loops in planar conformal $\mathcal{N}=4$ SYM theory and to all loop orders in QCD with massless quarks and at large $N_c$. For QCD, it reads
\begin{align}
\ln Q_s^2(L)=& \ Y+2\sqrt{4 b_0 Y}+3\xi_1(4b_0Y)^{1/6}\nn
&-\frac{1}{4}\left(3+8b_0\right)\ln Y+\mathcal{O}(1)\,,
\label{eq:rc-rhos-NLL-0} 
\end{align}
where $Y=\ln(L)$
and $b_0=1/\beta_0$ is the inverse of the one-loop coefficient of the QCD $\beta$-function. The sub-leading terms in the expansion \eqref{eq:rc-rhos-NLL-0} can be found in Eq.\,\eqref{eq:rc-rhos-NLL}.
It is quite remarkable that an exact, all order result can be obtained from perturbative QCD. In $\mathcal{N}=4$ SYM theory, the convergence of the perturbative series seems to be very fast, as we notice a very mild modification from the 2-loop to the 3-loop result. We observe the same feature in QCD: the corrections to the universal asymptotic behaviour beyond the DLA results \cite{Iancu:2014sha,Caucal:2022fhc} are small, and therefore, the DLA with running coupling turns out to be a very good approximation even at moderate values of $L$.

This paper is organized as follows: in the first section, we briefly review the calculation of the saturation momentum at tree-level and one-loop. We also set-up our notations for the rest of the paper. The second section discusses the non-linear quantum evolution of the dipole cross-section and the connection with the BFKL equation in the dilute, linear regime which drives the universality aspects of the saturation momentum for large system sizes. In section \ref{sec:as-fixed}, we revisit the fixed coupling evolution of the jet quenching parameter from the BFKL language, and address the corrections beyond the double logarithmic approximation in fixed coupling QCD and planar $\mathcal{N}=4$ SYM theory. Finally, in section \ref{sec:SL-all-order}, we solve the running coupling evolution at single logarithmic accuracy and present our all order result for the universal behaviour of $Q_s(L)$.

\section{The saturation momentum at tree-level and one loop}
\label{sec:Qs-tree}

The central object of this paper is the transverse momentum (TMB) distribution $\mathcal{P}(\kt,t)$ of a high energy parton in the colour representation $R=A,F$. It represents the probability to acquire a transverse momentum $\kt$ after a time $t$. We focus on the regime in which the momentum transfer from the medium is much smaller than the incoming parton energy $E=P^+$ such that the medium interactions do not alter significantly its direction of propagation. Under this approximation, the TMB distribution can be related to the forward scattering amplitude $\mathcal{S}(\rt)$ of an effective dipole with transverse size $\rt$ via a Fourier transform:
\begin{equation}
\mathcal{P}(\kt)=\int\dif^2\xt e^{-i\kt\cdot\rt}\mathcal{S}(\rt)\,.
\end{equation}
Assuming the interactions between the dipole and the medium scattering centers are local and instantaneous, with a collision rate $\mathcal{C}(\qt)$, the forward scattering amplitude exponentiates as follows
\begin{equation}
\mathcal{S}(\rt)=\exp\left[-\frac{C_R}{N_c}\sigma_\dip(\rt)L\right]\,,\label{eq:dipoleS}
\end{equation}
where $\sigma_\dip$ is the so-called dipole cross-section. At leading order, it is related to the collision rate $\mathcal{C}(\qt)$ according to
\begin{align}
\sigma_\dip(\rt)&\equiv \int\frac{\dif^2\qt}{(2\pi)^2}\left(1-e^{i\qt\cdot\rt}\right)\mathcal{C}(\qt)\,,\label{eq:sigmad-def}\,\\
             &=\frac{1}{4}\qhat(1/\rt^2,L)\, \rt^2(1+\mathcal{O}(\rt^2\mu^2))\,.\label{eq:qhat-def}
\end{align}
In the second equality, $\mu$ is a non-perturbative infrared momentum scale, typically of order of the plasma Debye mass $m_D\sim gT$ for a plasma at temperature $T$. Equation \eqref{eq:qhat-def} essentially defines the jet quenching parameter $\qhat(\kt^2,L)$ (in the adjoint representation) in the perturbative regime. At tree-level and for a static medium, $\qhat$ does not depend on the system size $L$. Applying the gradient $\nabla_{\rt}$ twice on Eq.\,\eqref{eq:sigmad-def}, and assuming that $\hat q$ is a weak function of $\rt$, it is straightforward to see that it can equivalently be defined as the second moment of the collision rate $\mathcal{C}(\qt)$, with an UV momentum cut-off set by $1/r_T$.

The saturation momentum is an emergent scale resulting from the unitarization  of the TMB distribution at small $k_T$ (but still much larger than $\mu$). It controls the transition between the dilute regime in which the TMB distribution has the typical Rutherford power law decay $\sim 1/\kt^4$ and the dense regime where the physics of multiple soft scatterings dominates, typically about $Q_s
\gg \mu$. This transition scale is defined by the implicit relation \cite{Kowalski:2003hm,Lappi:2011ju,Barata:2020rdn}
\begin{equation}
\mathcal{S}(1/Q_s^2(L))\equiv e^{-1/4}\Leftrightarrow \qhat(Q_s^2(L),L)L\equiv Q_s^2(L)\,.\label{eq:Qs-def}
\end{equation}
The number $e^{-1/4}$ is arbitrary here, and we will address the sensitivity of our results to this choice later in this paper. At tree-level, using the hard thermal loop (HTL) result for the collision rate $\mathcal{C}(\qt)$ that correctly accounts for modes with $|\qt|\ll T$, one finds that 
\begin{equation}
\qhat^{(0)}(\kt^2,L)=\qhat_0\ln(\kt^2/\mu^2)\,,\label{eq:qhat-tree}
\end{equation}
with $\qhat_0=\alpha_sN_cm_D^2T$ and $\mu=m_De^{-1+\gamma_E}/2$. The exact fixed coupling values for $\qhat_0$ and $\mu$ that also includes modes with $|\qt|\gtrsim T$ can be found in \cite{Arnold:2008vd,Caron-Huot:2008zna}. When using the one-loop running coupling the logarithmic dependence upon the hard scale $|\kt|\sim 1/|\rt|$ disappears resulting in $\qhat^{(0)}$ being a constant coefficient with $\qhat^{(0)}\propto\alpha_s(m_D^2)n$ \cite{Peshier:2006hi,Horowitz:2010yg,Kovchegov:2007vf}, where $n$ is the density of scattering centers. All these details will not matter in the following discussion, owing to the universal property of the asymptotic regime of $Q_s$. Also, note that $\qhat_0$ is proportional to $\alpha_s n$. Therefore, this "tree-level" computation is actually an all order resummation in the number of interactions with medium scattering centers. 
This should be kept in mind since when we will consider the weak coupling limit $\alpha_s\to 0$, it will be implicitly assumed that this limit is taken with the product $\alpha_s n L$ (that appears in the exponential Eq.\,\eqref{eq:dipoleS}) fixed. 

Using the expression \eqref{eq:qhat-tree}, we get the following result for the saturation momentum as a function of $L$:
\begin{equation}
Q_s^{2,(0)}(L)=\qhat_0L \ \textrm{W}_{-1}\left(-\frac{\mu^2}{\qhat_0L}\right)\simeq \qhat_0L\ln\left(\frac{\qhat_0L}{\mu^2}\right)\,,
\end{equation}
where $\mathrm{W}_p(x)$ is the Lambert function on the $p^{\rm th}$ branch. In what follows, it will be convenient to proceed with the following variables 
\begin{equation}
\rho_s(Y)=\ln(Q_s^2(L)/\mu^2)\,,\textrm{ }Y=\ln(L/\tau_0)\,,
\end{equation}
with $\tau_0\equiv \mu^2/\qhat_0$, the asymptotic behaviour of the saturation momentum at tree-level is given by
\begin{equation}
\rho_s^{(0)}(Y)=Y+\ln(Y)+...\label{eq:rhos-asymptotic-tree}
\end{equation}

At NLO, the TMB distribution, and therefore the saturation momentum itself are enhanced by large double logarithms $\alpha_s\ln^2(L/\tau_0)$ of the system size. Refs.~\cite{Liou:2013qya,Arnold:2021mow} report the following one loop result:
\begin{align}
&Q_s^{2}(L)=Q_s^{2,(0)}(L)\times \left[1+\bar\alpha_s\ln^2\left(\frac{L}{\tau_0}\right)\right.\nonumber\\
&\left.+\bar\alpha_s\left(2\ln(2)-\gamma_E-\frac{1}{3}\right)\ln\left(\frac{L}{\tau_0}\right)+\mathcal{O}(\alpha_s)\right]\,,\label{eq:Qs-NLO}
\end{align}
with $\bar \alpha_s=\alpha_sN_c/\pi$ and where the $\mathcal{O}(\alpha_s)$ denotes the $\alpha_s$ finite terms. $\gamma_E$ is the Euler-Mascheroni constant. The potentially large double logarithm in Eq.\,\eqref{eq:Qs-NLO} is the dominant radiative correction in the regime $E\gg\omega_c=\qhat L^2/2$ we are working, and needs to be resummed to all orders in perturbation theory when $\alpha_s Y^2=\mathcal{O}(1)$. 

\section{Non-linear evolution of the dipole cross-section}
\label{sec:qhat-evol}

A non-linear evolution equation resumming the double and single logarithms of Eq.\,\eqref{eq:Qs-NLO} has been proposed in \cite{Iancu:2014kga}. It is formulated directly in terms of the dipole cross-section $\sigma_\dip(\rt,\omega)$ which acquires a rapidity $\ln \omega $ dependence through the evolution, where $\omega\equiv k^+$ the light-cone energy of the gluon fluctuation. Schematically, this evolution equation reads
\begin{equation}
\frac{\partial \sigma_\dip}{\partial \ln \omega}=\mathcal{H}_{\qhat}\otimes \sigma_\dip(\rt,\omega)\,,\label{eq:non-linear-evol}
\end{equation}
where $\mathcal{H}_{\qhat}$ is a non-linear operator whose precise definition is not important to us (see Eq.\,(4.24) in \cite{Iancu:2014kga}). It satisfies the property that after one step of the evolution Eq.\,\eqref{eq:non-linear-evol}, one gets the double and single logarithmic terms in Eq.\,\eqref{eq:Qs-NLO}. 

Such evolution equation is difficult to solve both analytically and numerically. However, in this paper, we are mainly interested in the large $L$ limit of the saturation momentum $Q_s(L)$ arising from Eq.\,\eqref{eq:non-linear-evol}. In a series of recent papers, a new mathematical method based on the analogy between the evolution equation \eqref{eq:non-linear-evol} and the propagation of traveling wave fronts into unstable states has been developed in order to compute analytically the asymptotic behaviour of $Q_s(L)$, despite the absence of general analytic solutions of Eq.\,\eqref{eq:non-linear-evol}. In particular, the existence of traveling wave solutions to Eq.\,\eqref{eq:non-linear-evol} allows to simplify the problem and study the dilute (or linear) regime which drives the growth of the perturbations around the unstable state that determines the speed of the front $\dot\rho_s=\rmd\rho_s/\rmd Y$, in the presence of an absorptive boundary at $Q_s$. Essentially, the details of non-linear dynamics responsible for the saturation (unitarization) of the dipole S-matrix at $Q_s$ are irrelevant for the determination of the universal behaviour of the saturation scale \cite{2003,Munier:2009pc} (see also \cite{Mueller:2002zm,Munier:2003vc,Beuf:2010aw,Dominguez:2011gc,Dominguez:2011br} for concrete examples in the context of small $x$ physics). 

In order to define the dilute regime, we first recall that the formation time of the gluon fluctuation is of order $\tau\equiv 1/k^-=2\omega/\ktg^2$. In the dilute regime, the transverse momentum accumulated via multiple collisions over the formation time of the gluon, of order $\qhat\tau$, cannot exceed its transverse momentum. This criterion guarantees that the gluon is not sensitive to multiple soft scatterings over its formation time, so that only a single (or few \cite{Ghiglieri:2022gyv}, near the boundary of the constraint) scattering contributes to the cross-section.  Therefore, the dilute regime corresponds to 
\begin{equation}
 \ktg^2\gg Q_s^2(\tau)\sim \qhat \tau\,,\label{eq:dilute-regime}
\end{equation}
or, in terms of $\omega$, $\ktg^2\gg\sqrt{\qhat \omega}$.
In this regime, the non-linear evolution equation reduces to the well-known BFKL equation \cite{Iancu:2014kga}
\begin{align}
\frac{\partial \sigma_\dip(\rt,\omega)}{\partial \ln \omega}&=\frac{\abar}{2\pi}\int\rmd^2\zt \mathcal{K}_{r0z}\left[\sigma_\dip(\rt-\zt,\omega)\right.\nonumber\\
&\left.+\sigma_\dip(\zt,\omega)-\sigma_\dip(\rt,\omega)\right]\,,\label{eq:BFKL-cs}
\end{align}
with the LO BFKL kernel in coordinate space
\begin{equation}
\mathcal{K}_{xyz}=\frac{(\xt-\yt)^2}{(\zt-\xt)^2(\zt-\yt)^2}\,,
\end{equation}
and $\abar \equiv \alpha_s C_A/\pi$, albeit with an additional constraint that enforces the condition Eq.\,\eqref{eq:dilute-regime} in coordinate space (see for instance Eq.\,\eqref{eq:dilute-step} below).
We shall discuss the scale choice for the running coupling in the case of the jet quenching problem in the next section. In the BFKL equation in coordinate space \eqref{eq:BFKL-cs}, we have parametrically $|\kt|\sim 1/|\rt|$, (we remind that $\kt$ is the final transverse momentum of the incoming parton) and $|\ktg|\sim 1/|\zt|$ for the transverse momenta of the gluons along the ladder.

In addition to \cite{Iancu:2014kga}, the relevance of the BFKL equation in the renormalization of the jet quenching parameter has also been discussed from an effective field theory perspective in \cite{Vaidya:2021vxu,Vaidya:2021mly}. However, as we have anticipated it, the dynamics being of double logarithmic nature BFKL anf DGLAP evolutions are equally good to compute the asymptotics of the saturation scale for jet quenching. We stress that this is quite different in studies of proton structure at small-$x$ where the rapidity logarithm, $Y\equiv \ln 1/x $, is assumed to be much larger that the collinear logarithm  $\ln \kt^2 < \ln Q^2 \ll \ln 1/x$.  We will return to this crucial difference between small-$x$ and jet quenching evolution, and its implications, when we will discuss the BFKL kernel in Mellin space.

If one aims at reaching single logarithmic accuracy for $\sigma_\dip(\rt,\omega)$, one needs also to include the NLL BFKL evolution. This contribution is not manifest in the the fixed order computation in \cite{Liou:2013qya} because it appears at two loops in perturbation theory (it is a corrections of order $\alpha_s^2$). However, it is accompanied by a double log $\sim Y^2$, and therefore matters at single log accuracy since $\alpha_s^2Y^2=\mathcal{O}(1)$. 

It is convenient to write the full NLL BKFL equation using the variable $\rho=-\ln(\rt^2\mu^2)$ and $\eta=\ln(\omega/\omega_0)$ with the infrared energy scale $\omega_0=\qhat_0\tau_0^2/2$,
\begin{equation}
\frac{\partial \sigma_\dip}{\partial \eta}=\left[\abar\chi_{\rm LL}(-\partial_\rho)+\abar^2\chi_{\rm NLL}(-\partial_\rho)\right]\sigma_\dip(\rho,\eta)\,.\label{eq:NLL-BKFL-omega}
\end{equation}
In this equation, $\chi_{\rm LL}$ and $\chi_{\rm NLL}$ are the Mellin space representation of the BFKL leading log and next-to-leading log kernel. $\chi_{\rm LL}$ has the familiar expression
\begin{equation}
\chi_{\rm LL}(\gamma)=2\psi(1)-\psi(\gamma)-\psi(1-\gamma)\,,
\end{equation}
with $\psi(x)$ the digamma function. The expression for $\chi_{\rm NLL}$ is more complicated and can be found in \cite{Fadin:1998py}. In what follows, only the pole structure of $\chi_{\rm NLL}$ at $\gamma=1$ matters. Indeed, the jet quenching evolution problem is intrinsically double logarithmic in nature, in the sense that it is dominated by gluon fluctuations which are both strongly ordered in transverse momenta in the collinear regime $\kt^2\gg \ktg \gg ...\gg \mu^2$ and strongly ordered in energy or light-cone plus component $E\gg \omega \gg ...\gg \omega_0$. This double logarithmic regime is driven by the poles in $\gamma=1$ of $\chi_{\rm LL}$ and $\chi_{\rm NLL}$. In contrast, the standard small-$x$ evolution is driven by finite values of $\gamma$, i.e., $1/2$ and $0.372$  for BFKL with and without a saturation boundary, respectively \cite{Mueller:2002zm,Munier:2003vc,Munier:2003sj,kovchegov_levin_2012}. 

Before proceeding further we need first to address the issue of the large NLL correction in BFKL equation that are associated with a spurious triple pole in $\chi_{\rm NLL}$ at $\gamma=1$ due to the wrong choice of the evolution variable $\omega$ in the collinear regime \cite{Salam:1998tj,Ciafaloni:1999yw, Altarelli:2005ni,Beuf:2014uia}. The leading behaviour of $\chi_{\rm NLL}$ is indeed \cite{Fadin:1998py}
\begin{equation}
\chi_{\rm NLL}(\gamma)=-\frac{1}{(1-\gamma)^3}+\frac{B_g}{(1-\gamma)^2}+\mathcal{O}\left(\frac{1}{1-\gamma}\right)\,,\label{eq:chiNLL}
\end{equation}
with 
\begin{equation}
B_g=-\frac{11}{12}-\frac{N_f}{6N_c^3}\approx -\frac{11}{12}\,.
\end{equation}
In the second equality, we have used the large $N_c$ approximation.
The solution to this issue is well known. Instead of using $\omega=k^+$ as the evolution variable, one should use the lifetime $\tau=1/k^-=2\omega/\ktg^2$ of the gluon fluctuation. In terms of $Y=\ln(\tau/\tau_0)$, the NLL evolution equation for $\sigma_\dip(\rho,Y)$ is identical to Eq.\,\eqref{eq:NLL-BKFL-omega} (up to pure $\alpha_s$ corrections), but this time, $\chi_{\rm NLL}$ has no triple poles anymore \cite{Beuf:2014uia,Ducloue:2019ezk}. 
The equation we shall study is therefore 
\begin{equation}
\frac{\partial \sigma_\dip}{\partial Y}=\left[\abar\chi_{\rm LL}(-\partial_\rho)+\abar^2\tilde\chi_{\rm NLL}(-\partial_\rho)\right]\sigma_\dip(\rho,Y)\,.\label{eq:NLL-BKFL}
\end{equation}
with $\tilde\chi_{\rm NLL}=\chi_{\rm NLL}+1/(1-\gamma)^3+\mathcal{O}(1)$. Note that this equation has other issues in the double logarithmic \textit{anti-collinear} regime \cite{Ducloue:2019ezk}, but this regime is irrelevant in our case, as we shall see in the next section.

Finally, even though the dilute (linear) regime of the non-linear evolution drives the asymptotic behaviour of $\rho_s$, it is important to keep in mind that there is a major difference between BFKL evolution and the problem at hand. The latter pertains to the existence of a saturation (absorptive) boundary in the double logarithmic phase space. This saturation boundary can be accounted for via the following step function in coordinate space representation of the BFKL equation (using $1/\ktg^2\sim \textrm{max}((\xt-\zt)^2,\zt^2)$):
\begin{equation}
\Theta\left(\frac{4}{Q_s^2(\tau)}-\textrm{max}((\xt-\zt)^2,\zt^2)\right)\,,\label{eq:dilute-step}
\end{equation}
where $Q_s^2(\tau)= \qhat \tau$ is the saturation momentum evaluated at the gluon formation time $\tau\equiv1/k^-$. This step function enforces the transverse momentum of the gluon fluctuation to be larger than the saturation momentum so as the fluctuation is not affected by the LPM effect. 

It is worth noting that because the dynamics is dominated by strongly ordered transverse sizes in DLA, i.e., $\xt \ll \zt$ the theta function can be simplified as $ \Theta\left(4/Q_s^2(\tau)-\zt^2\right)$. With this simplification, one easily checks that Eq.\,\eqref{eq:BFKL-cs} with this constraint reduces to the non-linear evolution equation for $\qhat$ written in \cite{Blaizot:2014bha,Iancu:2014kga,Caucal:2021lgf} (see also Eq.\,\eqref{eq:DLA-eq-integral} below).

We are now left with the study of the NLL BFKL equation with a saturation boundary (or LO DGLAP as discussed in appendix \ref{app:dglap}). The exact implementation of this saturation condition is not decisive in the asymptotic regime, however, its very existence constrains the shape of the traveling wave ansatz that we shall discuss in what follows. 

\section{Revisiting the fixed coupling problem}
\label{sec:as-fixed}

In order to make the connection with our previous studies in \cite{Caucal:2021lgf,Caucal:2022fhc}, we shall consider the evolution of $\qhat(\rho,Y)$ instead of $\sigma_\dip(\rho,Y)$. The function $\qhat(\rho,Y)$ is defined according to Eq.\,\eqref{eq:qhat-def}, or equivalently in terms of $\rho$,
\begin{equation}
\qhat(\rho,Y)=4\mu^2\rme^{\rho}\sigma_\dip(\rho,Y)\,.
\end{equation}
Plugging this definition into Eq.\,\eqref{eq:NLL-BKFL}, using $\rme^{\rho}\chi(-\partial_\rho)e^{-\rho}=\chi(1-\partial_\rho)$ and the symmetry property $\chi(\gamma)=\chi(1-\gamma)$ of the kernel, one ends up with the following equation for $\qhat$
\begin{equation}
\frac{\partial \qhat (\rho,Y)}{\partial Y}=\left[\abar\chi_{\rm LL}(\partial_\rho)+\abar^2\tilde\chi_{\rm NLL}(\partial_\rho)\right]\qhat(\rho,Y)\,.\label{eq:NLL-BKFL-qhat}
\end{equation}
The saturation condition \eqn{eq:dilute-step}, that translates into $\rho > \rho_s(Y)$ in the new variables, is implicit in the above equation. 

The purpose of this section is  to first recover the known results \cite{Iancu:2014sha,Caucal:2022fhc}  for the asymptotic expansion of $\rho_s$ at DLA and fixed coupling from this equation and to discuss the case of the conformal $\mathcal{N}=4$ SYM theory in which the coupling does not run. We also demonstrate that contrary to the BFKL problem which is driven by the behaviour of the kernel around $\gamma_c\simeq 0.327$, the evolution of $\qhat$ is governed by the double logarithmic collinear regime and therefore driven by the behaviour of the kernel around $\gamma=0$ (or $\gamma=1$ for the dipole cross-section) as stated in the previous section.

At DLA and fixed coupling, one can neglect the NLL term and simply have 
\begin{equation}
\frac{\partial \qhat(\rho,Y)}{\partial Y}=\abar\chi_{\rm LL}(\partial_\rho)\qhat(\rho,Y)\,,\label{eq:LL-BKFL-qhat}
\end{equation}
for a fixed $\abar$.
The key starting point of our analysis consists in using the ansatz 
\begin{equation}
\qhat(\rho,Y)=e^{\rho_s(Y)-Y} e^{\beta x}f(x,Y)\,,\quad x=\rho-\rho_s(Y)
\end{equation}
in order to solve Eq.\,\eqref{eq:LL-BKFL-qhat} perturbatively in the limit $Y\to\infty$. It is motivated by the presence of the non-linear saturation condition in the evolution, $Q_s^2(L)=\qhat(Q_s^2(L),L)L$ which reads
\begin{equation}
\qhat(\rho_s(Y),Y)=e^{\rho_s(Y)-Y}
\end{equation}
in terms of $\rho_s$ and $Y$. Therefore, the function $f$ satisfies $f(0,Y)=1$ for all $Y$. This identity (more precisely, the $1$ on the r.h.s.) depends on the definition we adopt for $Q_s$. 

The form of this ansatz, though inspired by the study of traveling waves propagation governed by the Balitsky-Kovchekov (BK) equation \cite{Balitsky:1995ub, Kovchegov:1999yj} in small $x$ physics, differs from the latter in a crucial way. It involves an additional kinematic factor $ 1/\tau \sim \rme^{-Y} $ which is responsible for driving the evolution towards the double logarithmic regime $\gamma \sim 0+\cO(\abar^{1/2})$ instead of $\gamma_c \sim 0.327$. 
Nevertheless, we will see that our BFKL equation with saturation constraint admits traveling wave solutions (or geometric scaling solutions) of the form $\qhat(Y,\rho)=e^{\rho_s(Y)-Y}e^{\beta x}f(x)$ in the limit $Y\to \infty$. Since $x=\rho-\rho_s(Y)$, such solutions correspond indeed to the propagation of a front at the speed $\rmd \rho_s/\rmd Y$ from the left to the right along the $\rho$ axis.

As alluded to below Eq.\,\eqref{eq:Qs-def}, the saturation momentum is to be defined up to a undetermined multiplicative constant. This freedom can be absorbed into a redefinition of $Y$ through a constant shift (equivalent to a redefinition of the non-perturbative scale $\tau_0$). Such redefinition does not affect the universal terms in the asymptotic development of $\rho_s$ that we intend to calculate here, as can be checked by shifting $Y$ in the expressions \eqref{eq:rhos_nonlinear_fixed} and \eqref{eq:rc-rhos-NLL}.

From the existence of a scaling limit \cite{Caucal:2021lgf}, we expect the function $f(x,Y)$ to converge towards a well defined function $f(x)$ as $Y\to\infty$. The parameter $\beta$ will be determined later. Plugging this ansatz into Eq.\,\eqref{eq:LL-BKFL-qhat}, and expanding the kernel $\chi_{\rm LL}(\gamma)$ around $\gamma=\beta$, we find
\begin{equation}
(\dot\rho_s-1-\dot\rho_s\beta)f-\dot\rho_s \partial_x f+\partial_Y f=\abar\sum\limits_{p=0}^\infty\frac{\chi_{\rm LL}^{(p)}(\beta)}{p!}\partial_x^pf\,,\label{eq:fc-LL-exp}
\end{equation}
where
\beq
 \dot\rho_s (Y) \equiv \frac{\rmd \rho_s(Y)}{\rmd Y } \quad \text{and} \quad \chi_{\rm LL}^{(p)}(\beta) \equiv \frac{\del^p\chi_{\rm LL}(\beta) }{\del \beta^p} \,.
\eeq
Taking the limit $Y\to\infty$ and using the existence of a scaling limit for $f$, we end up with the following relations between the speed of the front 
\begin{equation}
c=\lim\limits_{Y\to\infty}\dot\rho_s(Y)\,,
\end{equation}
and the critical value $\beta_c$ of $\beta$ that minimizes \cite{Caucal:2021lgf} the velocity $c$:
\begin{align}
c-1-c\beta_c&=\abar\chi_{\rm LL}(\beta_c)\,,\\
-c&=\abar \chi_{\rm LL}'(\beta_c)\,,
\end{align}
by simply identifying the terms proportional to $f$ and $\partial_x f$.
This system cannot be solved analytically, however one can find the series expansion of $c$ and $\beta_c$ in powers of $\abar$:
\begin{align}
c&=1+2\sqrt{\abar}+2\abar+\mathcal{O}(\abar^{3/2})\,,\label{eq:c}\\
\beta_c&=\sqrt{\abar}-\abar+\mathcal{O}(\abar^{3/2})\,.\label{eq:betac}
\end{align}
At this stage, only the terms of order up to $\mathcal{O}(\sqrt{\abar})$ are under control since we neglect the NLL BFKL term in the evolution.
To recover the  DLA results reported in \cite{Caucal:2021lgf}, one can approximate $\chi_{\rm LL}$ by its most singular behaviour as $\gamma\to0$. The LL BFKL kernel behaves like
\begin{equation}
\chi_{\rm LL}(\gamma)=\frac{1}{\gamma}+2\zeta(3)\gamma^2+\mathcal{O}(\gamma^4)\,,
\end{equation}
at small $\gamma$, with $\zeta(x)$ the Riemann zeta function and $\zeta(3)\simeq 1.2$. Using $\chi_{\rm LL}=1/\gamma$, one can solve exactly the system \eqref{eq:c}-\eqref{eq:betac} (assuming $c\ge 1$) and find $
c = 1+2\sqrt{\abar+\abar^2}+2\abar$ and $\beta_c=\sqrt{\abar+\abar^2}-\abar$ in agreement with \cite{Caucal:2021lgf}.
The $\mathcal{O}(\sqrt{\alpha_s})$ and $\mathcal{O}(\alpha_s)$ terms are not affected by the use of the full BFKL kernel or its approximation $\chi_{\rm LL}\sim 1/\gamma$ (contrary to the BK case in small $x$ physics). One can actually check that the deviations enters at $\mathcal{O}(\alpha_s^2)$:
\begin{equation}
\beta_c=\abar^{1/2}-\abar+\frac{1}{2}\abar^{3/2}+2\zeta(3)\abar^2+\mathcal{O}(\abar^{5/2})\,.
\end{equation}
The last term cannot be obtained from the approximated kernel, as can be checked from the expansion of $\sqrt{\abar+\abar^2}-\abar$. It is a contribution beyond the double logarithmic regime of BFKL which would enter at the same order as the most singular term of the N$^3$LO BFKL kernel. This demonstrates that the asymptotic behaviour of $\dot\rho_s$ and $\qhat(\rho,Y)$ is mainly sensitive to the collinear double logarithmic regime of the BFKL evolution, since the details of the leading log kernel $\chi_{\rm LL}$ are not important up to order $\alpha_s^2$ (4-loops) in pQCD. In other words, the saddle point $\beta_c\sim\sqrt{\abar}$ is close to 0 for the $\qhat$ evolution, contrary to the case of small-$x$ evolution of the dipole operator where the saddle point lies at $\beta_c\approx 0.6275$. 

\begin{figure}[t]
 \centerline{\includegraphics[width=0.75\columnwidth]{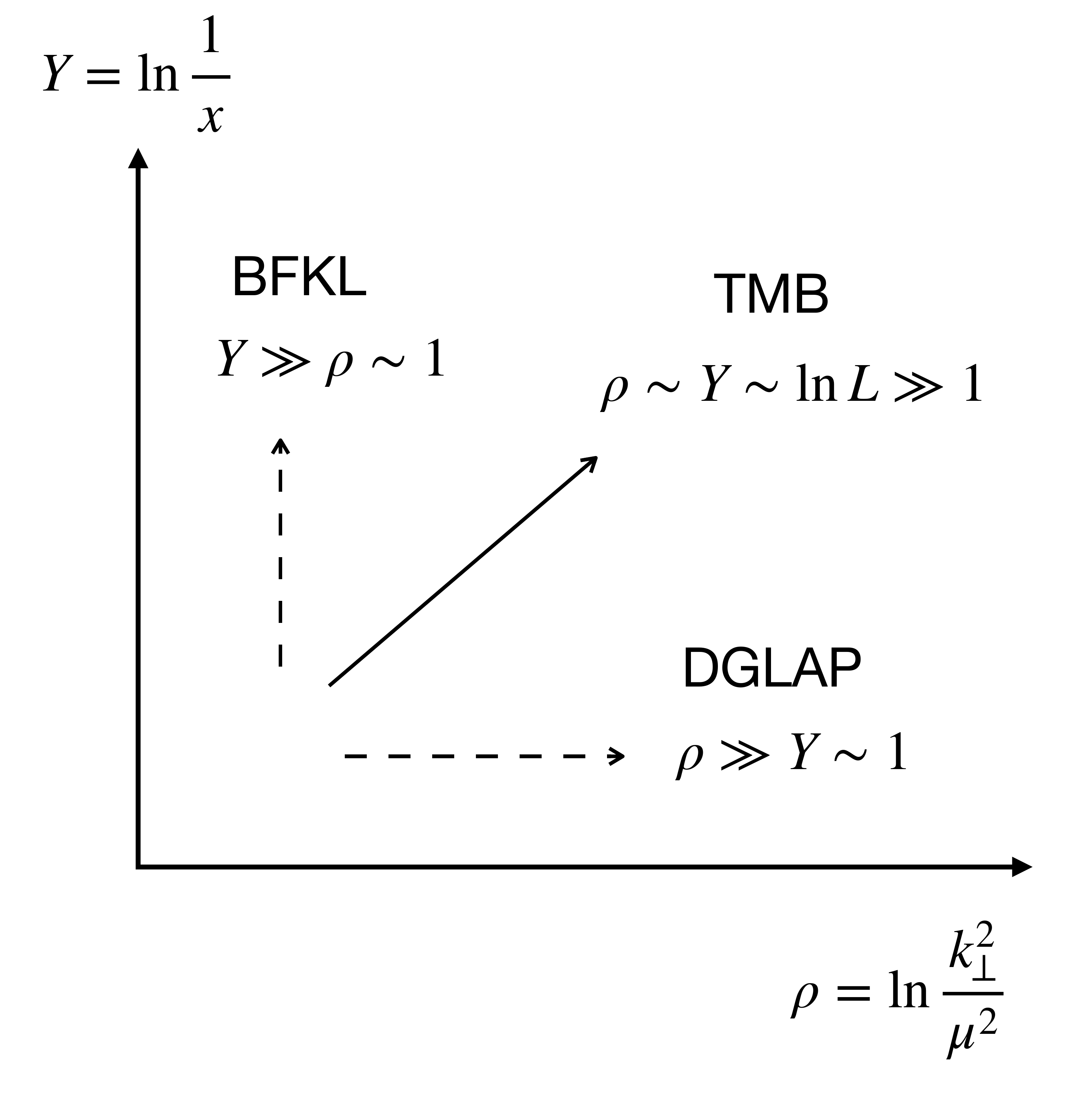}}
 \caption{The plane $(\rho,Y)$ and the standard DGLAP and BFKL regime of pQCD. The saturation scale $Q_s$ evolves along the diagonal of this diagram since $\rho\sim Y$.
 }\label{fig:DL-cartoon}
\end{figure}

Since the physics is dominated by the double logarithmic regime, which is common to both BFKL and DGLAP, one should be able to recover the results obtained in this paper from a BFKL approach using instead a DGLAP-like evolution with $\rho$ as the evolution variable. It turns out to be the case as shown in appendix \ref{app:dglap}. This is also illustrated in Fig.\,\ref{fig:DL-cartoon} where we display the $(\rho,Y)$ domain of pQCD as well as the standard DGLAP and BFKL directions along which the gluon distribution is evolved. The evolution of $Q_s$ follows the diagonal of this diagram as a consequence of the constraint $\rho\sim\rho_s\sim Y$, which is also where the DGLAP and BFKL evolution "merge" in the double logarithmic approximation. It is therefore natural that the corrections beyond DLA can be obtained from both DGLAP or BFKL approaches.

If one includes the NLL BFKL kernel $\tilde\chi_{\rm NLL}$, the system of equations \eqref{eq:c}-\eqref{eq:betac} is modified and the right hand side receives a contribution $\abar^2\tilde\chi_{\rm NLL}(\beta_c)$:
\begin{align}
c-1-c\beta_c&=\abar\chi_{\rm LL}(\beta_c)+\abar^2\tilde\chi_{\rm NLL}(\beta_c)\,,\label{eq:cNLL}\\
-c&=\abar \chi_{\rm LL}'(\beta_c)+\abar^2\tilde\chi_{\rm NLL}'(\beta_c)\,.\label{eq:betacNLL}
\end{align}
 With this additional term, we gain control over the terms of order $\mathcal{O}(\abar)$ in the $\alpha_s$ expansion of $c$ and $\beta_c$. Again, since the problem is dominated by the singular behaviour at $\gamma=0$, it is sufficient to use the approximation $\tilde\chi_{\rm NLL}(\gamma)=B_g/\gamma^2$, and one finds that 
\begin{align}
c&=1+2\sqrt{\abar}+(2+B_g)\abar+\mathcal{O}\left(\abar^{3/2}\right)\,.\label{eq:c-NLL-confQCD}
\end{align}
This result extends to single logarithmic accuracy the value of the traveling wave speed $c$. 

\begin{figure}[t]
 \centerline{\includegraphics[width=0.95\columnwidth]{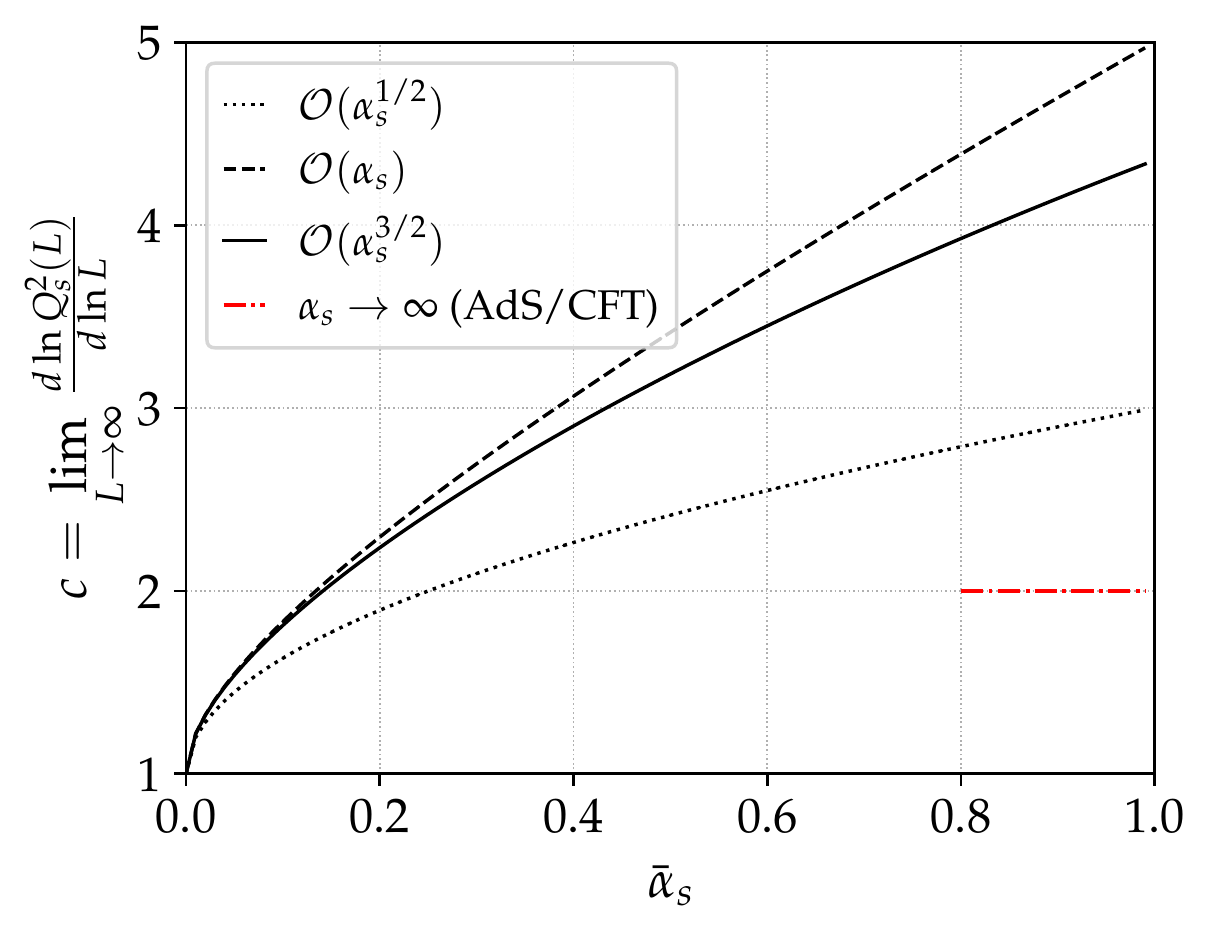}}
 \caption{The asymptotic limit of the front wave velocity as a function of the coupling constant $\abar=\alpha_sN_c/\pi$ in (planar) $\mathcal{N}=4$ SYM theory. 
 }\label{fig:c-N4SYM}
\end{figure}

The term of order $\mathcal{O}(\alpha_s^{3/2})$ receives contribution from both the sub-leading pole in $1/\gamma$ in the NLL BFKL kernel and from the pole in $1/\gamma^3$ of the N$^2$LL BFKL kernel. In fact, since the pole structure of the NLL and N$^2$LL BFKL equations are known thanks to the DGLAP/BFKL duality \cite{Marzani:2007gk,Costa:2012cb}, $\tilde\chi_{\rm NLL}\sim B_g/\gamma^2+a_{1,-1}/\gamma$ and $\tilde\chi_{\rm N^2LL}\sim a_{2,-3}/\gamma^3$, the value of $c$ can be known up to order $\alpha_s^{3/2}$, with the coefficient of the $\abar^{3/2}$ term equal to $1+3B_g-B_g^2+a_{1,-1}+a_{2,-3}$. This is a rather academic exercise given that the running of the coupling spoils this discussion, as we shall see in the next section. However, in $\mathcal{N}=4$ SYM theory, since the coupling does not run ($B_g=0$) and $a_{1,-1}=0$ \cite{Kotikov:2000pm}, we have the expansion
\begin{equation}
c= 1+2\sqrt{\abar}+2\abar+\left(1+a_{2,-3}^{\mathcal{N}=4}\right)\abar^{3/2}+\mathcal{O}(\abar^2)\,,
\end{equation}
with $a_{2,-3}^{\mathcal{N}=4}=-\zeta(2)=-\pi^2/6$ in the planar limit \cite{Gromov:2015vua,Velizhanin:2015xsa,Caron-Huot:2016tzz}.
The behaviour of this series is shown Fig.\,\ref{fig:c-N4SYM}, and compared to the strong coupling limit $c=2$ obtained from AdS/CFT \cite{Hatta:2007cs,Hatta:2007he,Dominguez:2008vd}. 

It would be interesting to exploit the BFKL/DGLAP duality (or existing results for N$^p$LL BFKL equation with $p\ge 4$ \cite{Velizhanin:2021bdh}) in order to extend this calculation to higher orders in $\alpha_s$ and see the convergence of the series and the approach towards the strong coupling regime.

Turning back to QCD, one may want to understand physically where the coefficient of the $\alpha_s$ term in $c$ given by Eq.\,\eqref{eq:c-NLL-confQCD} comes from. The $B_g$ term is the finite part of the gluon splitting function, and therefore this contribution is associated with collinear but non-soft splittings in the evolution. On the other hand, the $2$ term seems more mysterious at first glance. We argue that it comes from the feedback of the quantum evolution of $Q_s$ on the evolution of the dipole cross-section $\sigma_\dip$ (recall that these physical quantities are related to one another by the non-linearity of the evolution). Indeed, the fixed coupling DLA equation with $\chi_{\rm LL}=1/\gamma$ is equivalent to the equation
\begin{equation}
\frac{\partial \qhat}{\partial Y}=\abar \int^\rho\rmd\rho' \ \qhat(\rho',Y)\,,\label{eq:DLA-eq-integral}
\end{equation}
owing to the formal relation $\int^\rho\rmd\rho'=1/\partial_\rho$.
In this equation, the lower bound is set by the single scattering criterion $\ktg^2\ge Q_s^2$ i.e.\ $\rho'>\rho_s(Y)$. If we neglect quantum evolution in this lower bound and use the tree-level "classical" expression  $\rho_s(Y)=Y$ instead, to constraint the $\rho'$ integral such that $\rho' > Y$, this equation can be solved exactly. Defining $Q_s^2(L)=\qhat(\qhat_0 L,L)L$ in agreement with the previous approximation, the resummed value of the saturation momentum is given by \cite{Iancu:2014sha,Mueller:2016xoc}
\begin{equation}
Q_s^2(L)=Q_s^{2,(0)}(L)\times \frac{\textrm{I}_1\left(2\sqrt{\abar Y^2}\right)}{\sqrt{\abar Y^2}}\,,\label{eq:Qs-analytic-linear}
\end{equation}
for a constant initial condition (hence $Q_s^{2,(0)}(L)=\qhat_0 L$ here). Here $\textrm{I}_n(x)$ is the modified Bessel function of rank $n$. This result shows very clearly the double logarithmic resummation structure, since the saturation momentum is expressed as a function of $\alpha_s Y^2=\alpha_s\ln^2(L/\tau_0)$. In the asymptotic limit, assuming the strong condition $\alpha_sY^2\gg 1$ (instead of $\alpha_sY^2\sim 1$ only), one can easily derive the behaviour of the saturation scale from Eq.\,\eqref{eq:Qs-analytic-linear}:
\begin{equation}
\rho_s(Y)=(1+2\sqrt{\abar})Y-\frac{3}{2}\ln(Y)+\mathcal{O}(1)\,.\label{eq:rhos_linear_fixed}
\end{equation}
In DLA evolution with linearization of the saturation boundary, one observes that the traveling wave speed is $c=1+2\sqrt{\abar}$ without terms of order $\alpha_s$. Hence, one can interpret the 2 term in the $\abar$ coefficient of Eq.\,\eqref{eq:c-NLL-confQCD} as a single log effect coming from the quantum evolution of $Q_s$ itself. A similar argument is presented in \cite{Iancu:2014sha}.

In Eq.\,\eqref{eq:rhos_linear_fixed}, we also show the sub-asymptotic correction to the constant speed motion of the saturation front in the case of the linearized DLA evolution. For the full NLL problem, the sub-asymptotic corrections to $\dot\rho_s$ can be obtained using the same method as in \cite{Caucal:2021lgf,Caucal:2022fhc}. Namely, we expand $f(x,Y)$ as an infinite series in powers of $Y^{-1/2}$ multiplied by diffusive scaling functions $G_n$ \cite{Brunet:1997zz,2000,Munier:2003sj}:
\begin{equation}
f(x,Y)=\sum\limits_{n=-1}^{\infty}Y^{-n/2}G_n\left(\frac{x}{Y^{1/2}}\right)\,.\label{eq:fc-leadingedge}
\end{equation}
Using this expansion and solving order by order for the functions $G_n$ with appropriate boundary conditions, we can fix the asymptotic development of $\dot\rho_s$. This is shown in appendix \ref{app:Gcalc}. We find that this development can be expressed in terms of $c$ and $\beta_c$ as
\begin{align}
\dot\rho_s&=c+\frac{3}{2(\beta_c-1)}\frac{1}{Y}+\frac{3}{2(\beta_c-1)^2}\sqrt{\frac{2\pi}{\chi''(\beta_c)}}\frac{1}{Y^{3/2}}\nonumber\\
&+\mathcal{O}\left(Y^{-2}\right)\,,\label{eq:rhos_nonlinear_fixed}
\end{align}
with $ \chi''(\beta_c)=\abar \chi_{\rm LL}''(\beta_c)+\abar^2\tilde\chi_{\rm NLL}''(\beta_c)+...$.
This expression encompasses all universal terms in the expansion of $\dot\rho_s$ at fixed coupling. The first non-universal terms that are sensitive to the initial condition appear at the order $Y^{-2}$. Such terms can be easily produced by shifting the value of $Y$ by a constant to absorb a change in the non-perturbative parameters $\tau_0$ or $\mu$.

When using the DLA, i.e.\ $\chi=\abar/\gamma$, one recovers from \eqn{eq:rhos_nonlinear_fixed} the expression found in \cite{Caucal:2021lgf,Caucal:2022fhc}. 

Each coefficient of the $Y$ powers has an $\alpha_s$ expansion which can be obtained from the one of $c$ and the shape of the BFKL kernels near the double logarithmic regime $\gamma=0$. These expansions are given in appendix~\ref{app:rhos-exp} up to order $\alpha_s^{3/2}$ in $\mathcal{N}=4$ SYM theory. They are systemically improvable by including higher logarithmic orders in the BFKL equation. As we shall see in the next section, conformal symmetry breaking in QCD changes this simple picture in a quite dramatic way: the coefficients of the universal terms in the $\rho_s$ expansion are entirely fixed by the BFKL kernel at NLL order.

\begin{figure}[t]
 \centerline{\includegraphics[width=0.95\columnwidth,page=2]{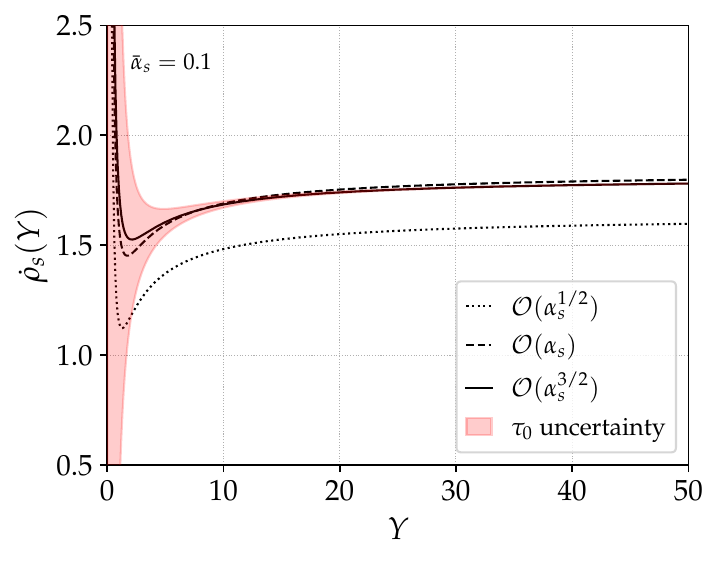}}
 \caption{The front wave velocity as a function of $Y$ in (planar) $\mathcal{N}=4$ SYM theory with $\abar=0.1$. The red band shows the uncertainty related to the unknown $Y^{-2}$ corrections in $\dot\rho_s$, obtained by varying $\tau_0$ by a factor of $2$ around a central value. For $Y\gtrsim 5$, next-to-double log corrections dominate over the non-universal sub-asymptotic corrections in $Y^{-2}$.
 }\label{fig:rhosY-N4SYM}
\end{figure}

The expansion \eqref{eq:rhos_nonlinear_fixed} of $\dot\rho_s(Y)$ as a function of $Y$ is shown in Fig.\,\ref{fig:rhosY-N4SYM} in the case of $\mathcal{N}=4$ SYM theory, at DLA, order $\alpha_s$ and order $\alpha_s^{3/2}$ for $\abar=0.1$. The convergence is very good, and one notices that the corrections beyond the double logarithms are more important than the non-universal sub-asymptotic $1/Y^2$ correction (estimated by varying $\tau_0$ by factors between 0.5 and 2) for $Y\gtrsim 4$.

\section{Running coupling effects and all-order result for the saturation momentum}
\label{sec:SL-all-order}

We now address the running coupling problem and compute the universal asymptotic expansion of $\dot\rho_s$ in QCD at single logarithmic accuracy. To do so, it is sufficient to consider the one-loop running coupling:
\begin{equation}
\alpha_s(\rho)=\frac{b_0}{\rho+\rho_0}\,,
\end{equation}
with $b_0=1/\beta_0=-1/B_g$ and $\rho_0=\ln(\mu^2/\Lambda_{\rm QCD}^2)$.
When including running coupling effects, one has to be careful with the prescription for the running scale. At DLA, it has been argued in \cite{Liou:2013qya,Iancu:2014sha} that the running coupling should be implemented in the following way:
\begin{equation}
\frac{\partial \qhat}{\partial Y}= \int^\rho\rmd\rho' \ \abar(\rho')\qhat(\rho',Y)\,,
\end{equation}
so that the standard DGLAP equation in the DLA is recovered (cf appendix~\ref{app:dglap}).
In Mellin space, the corresponding NLL equation is 
\begin{equation}
\frac{\partial \qhat}{\partial Y}=\chi_{\rm LL}(\partial_\rho)\left[\abar(\rho)\qhat(\rho,Y)\right]+\abar^2(\rho)\tilde\chi_{\rm NLL}(\partial_\rho)\qhat(\rho,Y)\,,\label{eq:NLL-BKFL-qhat-rc}
\end{equation}
which is slightly different from the running coupling BFKL equation where $\abar(\rho)$ is outside of the BFKL kernel. 
In fact, these two prescriptions, or scheme choices, are related to one another by a modification of the NLL kernel \cite{Marzani:2007gk}. For the NLL term proportional to $\abar^2$, the differences between these two schemes is of higher order (NNLL), which explains why we have written $\abar^2(\rho)$ outside of the NLL kernel.

We therefore need to study the evolution equation \eqref{eq:NLL-BKFL-qhat-rc} with the pole structure of the NLL BFKL kernel given by Eq.\,\eqref{eq:chiNLL}. As shown in \cite{Caucal:2022fhc}, the DLA running coupling involves a modified geometric scaling limit which takes the form
\begin{equation}
\qhat(\rho,Y)=e^{\rho_s(Y)-Y}e^{\beta x}f(x,Y)\,,\quad x=\frac{\rho-\rho_s(Y)}{\sqrt{Y}}\,,
\end{equation}
with the function $f(x,Y)$ having a scaling limit $f(x)$ as $Y\to\infty$.
We then plug this ansatz inside the evolution equation and expand the kernels $\chi_{\rm LL}$ and $\chi_{\rm NLL}$ around the DLA "saddle point" $\beta/\sqrt{Y}$, as expected from the fixed coupling result where $\beta_c\sim\sqrt{\alpha_s(\rho_s)}\propto Y^{-1/2}$. After these manipulations, we find the equation
\begin{align}
&\left(\dot\rho_s-1-\frac{\beta x}{2Y}-\frac{\beta\dot\rho_s}{Y^{1/2}}\right)f-\left(\frac{x}{2Y}+\frac{\dot\rho_s}{Y^{1/2}}\right)\frac{\partial f}{\partial x}+\frac{\partial f}{\partial Y}\nonumber\\
&=\sum\limits_{p=0}^\infty\frac{\chi_{\rm LL}^{(p)}(\beta/Y^{1/2})}{p!Y^{p/2}}\partial_x^p \left[\frac{b_0}{\rho_s+xY^{1/2}+\rho_0} f\right]\nonumber\\
&+\left(\frac{b_0}{\rho_s+xY^{1/2}+\rho_0}\right)^2\sum\limits_{p=0}^\infty\frac{\tilde\chi_{\rm NLL}^{(p)}(\beta/Y^{1/2})}{p!Y^{p/2}}\partial_x^p f \,.\label{eq:basic-equation-rc}
\end{align}
As in the fixed coupling problem, let us first determine the exact location of the saddle point proportional to $\beta$ and the value of the first non trivial correction to $\dot\rho_s=1$ such that 
\begin{equation}
\dot\rho_s=1+\frac{c}{Y^{1/2}}+...
\end{equation}
This behaviour is dictated by the homogeneity of Eq.\,\eqref{eq:basic-equation-rc}. Seeking for the leading power in $1/\sqrt{Y}$, one gets the two following equations
\begin{align}
c-\beta_c&=b_0/\beta_c\,,\\
-1&=-b_0/\beta_c^2\,,
\end{align}
after identification of the coefficients in front of $f$ and $\partial_x f$.
Therefore, one recovers the scaling limit of the running coupling evolution equation at DLA:
\begin{align}
c&=2\sqrt{b_0}\,,\\
\beta_c&=\sqrt{b_0}\,.
\end{align}
We emphasize that these relations are exact to all orders in pQCD. Indeed, contrary to the fixed coupling problem, the NLL kernel (and the higher orders) do not determine the value of $c$ and $\beta_c$, thanks to the asymptotic freedom property of QCD which imposes $\alpha_s$ to decay as $1/Y$ at large $Y$.

To obtain the sub-asymptotic corrections, we use again the leading edge expansion
\begin{equation}
f(x,Y)=\sum\limits_{n=-1}^{\infty}Y^{-n/6}G_n\left(\frac{x}{Y^{1/6}}\right)\,,\label{eq:rc-leading edge}
\end{equation}
with a $1/6$ diffusive exponent \cite{Caucal:2022fhc}. The interplay between this expansion and the asymptotic series of $\dot\rho_s$ enables to compute the sub-asymptotic corrections. Namely, the determination of the function $G_{-1}$ fixes the correction of order $Y^{-5/6}$ in $\dot\rho_s$, the function $G_0$ determines the correction of order $Y^{-1}$, and so forth (see appendix~\ref{app:Gcalc}). After a straightforward calculation of these functions similar to the one in \cite{Caucal:2022fhc}, we find for the universal asymptotic series,
\begin{align}
\dot\rho_s&=1+\frac{4b_0}{\bar Y^{1/2}}+\frac{2\xi_1b_0}{\bar Y^{5/6}}+b_0\left(1-8b_0+4b_0B_g\right)\frac{1}{\bar Y}\nonumber\\
&-\frac{7\xi_1^2b_0}{270}\frac{1}{\bar Y^{7/6}}
-(5+1944b_0)\frac{\xi_1b_0}{81}\frac{1}{\bar Y^{4/3}}\nonumber\\
&-2b_0^2\left(1-8b_0+4b_0B_g\right)\frac{\ln(\bar Y)}{\bar Y^{3/2}}+\mathcal{O}(\bar Y^{-3/2})\,,\label{eq:rc-rhos-NLL}
\end{align}
with $\bar Y = 4b_0Y$ and $\xi_1\simeq -2.338$ is the rightmost zero of the Airy function. Setting $B_g=0$ in this expression, we recover the DLA result obtained in \cite{Caucal:2022fhc}. Note also that, since $B_g=-1/b_0$, the coefficient of the $1/\bar Y$ term is equal to $-b_0(3+8b_0)$.
Surprisingly, the $\mathcal{O}(Y^{-7/6})$ and $\mathcal{O}(Y^{-4/3})$ terms are not affected by the single log corrections (they do not depend on $B_g$). The NLL DGLAP contribution (which appears as a $1/\gamma$ pole in the NLL BFKL kernel \cite{Marzani:2007gk}) starts contributing at order $Y^{-3/2}$, but this order is non-universal. The $2\zeta(3)\gamma^2$ term in $\chi_{\rm LL}$ starts contributing at order $n=6$ in the leading edge expansion, meaning that the correction of order $Y^{-2}$ in $\dot\rho_s$ would depend on $\zeta(3)$.  

It is also interesting to compare with the universal asymptotic expansion of the saturation scale in the case of small-$x$ evolution (see \cite{gribov1983semihard,Iancu:2002tr,Mueller:2002zm,Munier:2003sj,Beuf:2010aw}). For the energy dependence of $Q_s$, the universal terms also depend only on the LL and NLL BFKL evolution, but the coefficients of the development have a much stronger dependence on the shape of the LL and NLL BFKL kernel since they depend on the first five derivatives of the kernel at the small-$x$ saddle point $\beta_c=0.6275$. Again, this is a specificity of the jet quenching problem, which is controlled by double logarithmic physics.

To sum up, the effect of the running coupling is to reduce the sensitivity of the sub-asymptotic corrections to higher orders in the resummmation. Eventually, the universal asymptotic expansion of the saturation momentum is entirely given by the leading poles of the LL+NLL BFKL kernel in Mellin space or the singular plus finite part of the LO DGLAP splitting function. This is a consequence of the double logarithmic nature of the problem at hand.

\begin{figure}[t]
 \centerline{\includegraphics[width=0.95\columnwidth,page=2]{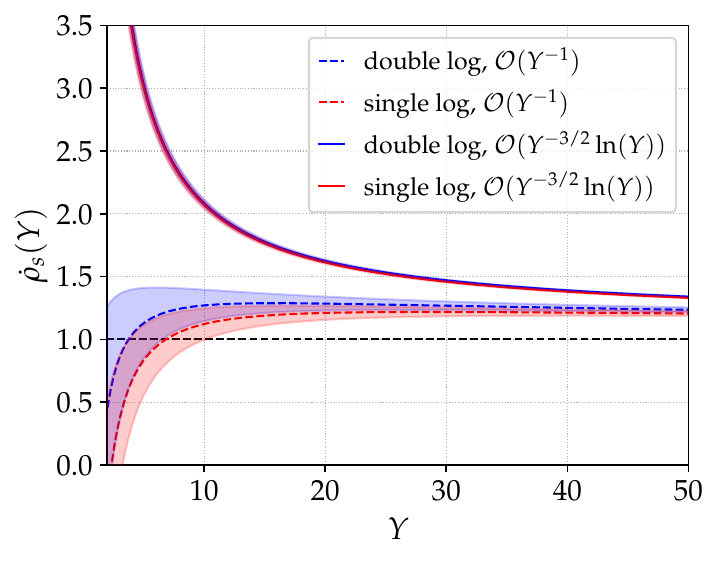}}
 \caption{The front wave velocity as a function of $Y$ in QCD. The bands show the uncertainty related to the $\bar Y^{-7/6}$ (dotted) and $\bar Y^{-3/2}$ (full) corrections in $\dot\rho_s$, obtained by adding the corresponding power multiplied by a coefficient $\kappa b_0$ with $|\kappa|\le 10$. This comprehensive interval for $\kappa$ is due to the potentially large non-universal $Y^{-3/2}$ coefficient, as noted in \cite{Caucal:2022fhc}.
 }\label{fig:dotrhos-QCD}
\end{figure}

A comparison between the double and single log result is shown Fig.\,\ref{fig:dotrhos-QCD} for two truncations of the asymptotic expansion, either up to the $1/Y$ term or up to the $\ln(Y)/Y^{3/2}$. In the latter case, one observes that the effect of single log corrections is very mild. The band show the estimated uncertainty coming from the following sub-leading correction in the large $Y$ development of $\dot\rho_s$. As expected, pushing the series up to order $1/Y^{3/2}$ reduces this source of uncertainty, and the corrections from single-log effects fall within the bands. However, we also know that the asymptotic series converges very slowly at small $Y \lesssim 10$ when using the non-linear saturation boundary $\rho_s(Y)$ \cite{Caucal:2022fhc}, so our estimation of the uncertainty coming from the non-universal $Y^{-3/2}$ term should be taken with a grain a salt in that domain.

\section{Summary and outlook}

In this paper, we have computed for the first time the universal behaviour --- independent of the non-perturbative, tree-level physics --- of the saturation momentum associated with transverse momentum broadening of high energy partons in QCD media beyond the double logarithmic approximation. Our study relies on two pillars (i) the universal terms of the asymptotic series of $Q_s(L)$ at large $L$ are essentially controlled by the linearized evolution equation for the dipole cross-section. This is a consequence of the mathematical mapping between this evolution equation and equations describing the propagation of traveling wave fronts into unstable states (ii) the quantum evolution of the dipole cross-section in the jet quenching problem is dominated by the double logarithmic regime of pQCD since for $\kt^2\sim Q_s^2$, $\ln(\kt^2/\mu^2)\sim \ln(L/\tau_0)\sim \ln(1/x)$. As a consequence, the universal asymptotic series can be obtained either from a BFKL or DGLAP approach, and the coefficient of the series are only sensitive to the shape of these two kernels close to the double logarithmic regime.

For a fixed coupling evolution, as in the case of the supersymmetric $\mathcal{N}=4$ SYM theory, we have obtained the coefficients of the development of $\dot\rho_s$ up to 3-loops order $\alpha_s^{3/2}$, thanks to the known pole structure of BFKL at three loops in this theory. Using DGLAP/BFKL duality, we believe that our method can be straightforwardly extended to higher orders. As we observe a good convergence of the coefficients for $\alpha_s$ up to $\sim 0.4$, we have not tried to extend further our calculation.

The running coupling dramatically changes both the behaviour of the large $L$ expansion and the way higher order corrections in the resummation appear. Interestingly enough, the universal terms depend only on the one-loop QCD $\beta$ function.
  The main result of this paper, the asymptotic behaviour of $\rmd\rho_s/\rmd Y$ in QCD given by Eq.\,\eqref{eq:rc-rhos-NLL}, is therefore exact to all orders in perturbation theory. At large $L$, the effect of corrections beyond the double logarithmic approximation turns out to be very mild, demonstrating the excellent convergence of $\rho_s$ after quantum evolution.

The main limitation of our work is the large $L$ assumption. Despite the calculation of the sub-asymptotic corrections which, in principle, enable to reach phenomenological values for the system size $L$, our work should be supplemented by a study of the moderate and small $Y$ domain, which is not anymore driven by universality. This requires to solve numerically a non-linear evolution for the jet quenching parameter $\qhat$, valid at single logarithmic accuracy, along the lines of \cite{Iancu:2014kga}. This is a very challenging task in practice, that we leave for future works. An other possibility would be to take advantage of recent progress in quantum computing to address this problem, following the approach of \cite{Li:2021zaw,Barata:2022wim}, in which radiative corrections are straightforward to include.

\smallskip
\noindent{\bf Acknowledgements.}
We thank Edmond Iancu for insightful discussions about the strong coupling limit of the saturation scale.
 This work was supported by the U.S. Department of Energy, Office of Science, Office of Nuclear Physics, under contract No. DE- SC0012704. Y. M.-T. acknowledges support from the RHIC Physics Fellow Program of the RIKEN BNL Research Center.
 
\bibliographystyle{apsrev4-1}
\bibliography{biblio}

\appendix

\section{Single logarithmic corrections from DGLAP-like evolution}
\label{app:dglap}

In this appendix, we compute the single log corrections from a DGLAP evolution. Our starting point is the DGLAP equation for the gluon distribution function $xg(x,Q^2)$:
\begin{equation}
\frac{\partial x g(x,Q^2)}{\partial \ln(Q^2)}=\frac{\alpha_s(Q^2)}{2\pi}\int_x^1\frac{\rmd z}{z}P_{g}(z)xg(x/z,Q^2)\,.
\end{equation}
In the double logarithmic approximation, one uses $P_{g}(z)=2C_A/z$, giving
\begin{equation}
\frac{\partial x g(x,Q^2)}{\partial \ln(Q^2)}=\abar\int_x^1\frac{\rmd z}{z}\frac{x}{z}g(x/z,Q^2)\,.
\end{equation}
To obtain the corresponding equation for $\qhat$, we use the fact that in the dilute limit, $\qhat(\rho,Y)\propto xg(x,Q^2)$ \cite{Caucal:2022fhc}, with the identification $Y=-\ln(x)$ and $\rho=\ln(Q^2/\mu^2)$, so that the DGLAP evolution of $\qhat$ reads
\begin{equation}
\frac{\partial \qhat(Y,\rho)}{\partial \rho}=\abar(\rho)\int_0^Y\rmd Y'\qhat(Y',\rho)\,,
\end{equation}
which is almost equivalent to the non-linear equation studied in \cite{Caucal:2021lgf} modulo the replacement $Y\to\mathrm{min}(Y,Y_s(\rho))$ in the upper limit of the $Y'$ integral (with $Y_s(\rho)$ the inverse function of $\rho_s(Y)$).
At single log accuracy, one can use the following approximation of $P_{g}(z)$:
\begin{equation}
P_{g}(z)=\frac{2C_A}{z}\left(1+B_g z\right)\,,
\end{equation}
with $B_g=-11/12-N_f/(6N_c^3)$ the finite part of the $g\to gg$ (and $g\to q\bar q$) splitting function.
Hence, the equation we shall study is
\begin{equation}
\frac{\partial \qhat(Y,\rho)}{\partial \rho}=\abar(\rho)\int_0^{Y}\rmd Y'\,\left[1+B_ge^{Y'-Y}\right]\qhat(Y',\rho)\,.
\end{equation}
The equation above can be written in a fully differential form:
\begin{align}
\frac{\partial^3\qhat(Y,\rho)}{\partial Y^2\partial\rho}&+\frac{\partial^2\qhat(Y,\rho)}{\partial Y\partial \rho}-\abar(\rho)(1+B_g)\frac{\partial \qhat(Y,\rho)}{\partial Y}\nonumber\\
&-\abar(\rho)\qhat(Y,\rho)=0\,.\label{eq:dglap-SL}
\end{align}
Let us consider the fixed coupling approximation $\abar(\rho)=\abar$, and look for an asymptotic solution of the form
\begin{equation}
\qhat(Y,\rho)=e^{\rho_s(Y)-Y}e^{\beta x}\,,\quad x=\rho-\rho_s(Y)\,,
\end{equation}
with $\dot\rho_s= c$.
We have 
\begin{equation}
\frac{\partial^2\qhat(Y,\rho)}{\partial Y\partial \rho}=e^{\beta x}\left[(c-1)\beta-c\beta^2\right]\,.
\end{equation}
Plugging our ansatz inside the differential equation, one finds the following relation between $\beta$ and $c$:
\begin{equation}
(\beta-1)(1+c(\beta-1))\beta c+\abar(B_g+(\beta-1)(1+B_g)c)=0\,.
\end{equation}
The extremum condition, obtained by differentiating the relation above with respect to\ $\beta$ gives
\begin{equation}
-1+\abar(1+B_g)+2\beta+c-4\beta c+3\beta^2c=0\,.
\end{equation}
This system can be exactly solved, but it is more enlightening to find the $\abar$ expansion of $c$ and $\beta$. 
The following two developments are solution
\begin{align}
c&=1+2\sqrt{\abar}+(2+B_g)\abar+\mathcal{O}(\abar^{3/2})\,,\\
\beta&=\sqrt{\abar}+(B_g-1)\abar+\mathcal{O}(\abar^{3/2})\,.
\end{align}
This is the same result as the one obtained from the fixed coupling NLL BFKL evolution.

\section{Leading edge expansion calculation}
\label{app:Gcalc}

We briefly reproduce here the calculation of the universal terms in the leading edge expansion detailed in \cite{Caucal:2022fhc}. We essentially detail the calculation of the first term $G_{-1}(z)$ in this series, which fixes the first sub-asymptotic correction to $\dot\rho_s$. The computation of the higher order terms proceeds in a similar fashion.

\paragraph{Fixed coupling.} We first insert Eq.\,\eqref{eq:fc-leadingedge} in Eq.\,\eqref{eq:fc-LL-exp}, with the asymptotic series
\begin{equation}
\dot\rho_s=c+\frac{\delta_1}{Y}+...\,,
\end{equation}
Gathering the leading terms proportional to $1/Y$ in the resulting equation, we get the following differential equation satisfied by the function $G_{-1}$:
\begin{equation}
-\frac{1}{2}\chi''(\beta_c)G_{-1}''-\frac{1}{2}zG_{-1}'+\left(\frac{1}{2}+\delta_1-\delta_1\beta_c\right)G_{-1}=0\,.
\end{equation}
This equation can be solved analytically with the initial conditions $G_{-1}(z)=\textrm{cste}\times z+\mathcal{O}(z^2)$. This initial condition comes from the saturation constraint on the evolution equation which imposes $f(0,Y)=1$ and therefore $G_{-1}(0)=0$. Demanding the solution to decay at large $z$ and to be positive, one can further constrain the value of $\delta_1$ to be 
\begin{equation}
\delta_1=\frac{3}{2(1-\beta_c)}\,,
\end{equation}
so that the function $G_{-1}$ reads
\begin{equation}
G_{-1}(z)=\textrm{cste}\times \beta_c z \exp\left(-\frac{z^2}{2\chi''(\beta_c)}\right)\,.
\end{equation}

\paragraph{Running coupling.} The running coupling case is very similar, the only difference comes from the evolution equation \eqref{eq:basic-equation-rc} which imposes a diffusive power $1/6$ in the leading edge expansion. The homogeneity of Eq.\,\eqref{eq:basic-equation-rc} constrain the possible power of the sub-leading asymptotic corrections to $\dot\rho_s$. We have in particular:
\begin{equation}
\dot\rho_s=1+\frac{c}{Y^{1/2}}+\frac{\delta_1}{Y^{5/6}}+...\,,
\end{equation}
  Plugging the series \eqref{eq:rc-leading edge} and this development for $\dot\rho_s$ inside Eq.\,\eqref{eq:basic-equation-rc} and expanding in powers of $Y$ the result, we find the equation 
\begin{equation}
-G_{-1}''(z)+\left(\frac{1}{2}b_0z+\delta_1\sqrt{b_0}\right)G_{-1}(z)=0\,,
\end{equation}
by identifying the $Y^{-2/3}$ power. The solution to this equation with boundary conditions $G_{-1}(z)\propto z$ at small $z$ and $G_{-1}(z)\to 0$ at large $z$ is (we denote $\textrm{Ai}$ the Airy function of the first kind)
\begin{equation}
G_{-1}(z)=\textrm{cste}\times \textrm{Ai}\left(\xi_1+2^{-1/3}b_0^{1/3}z\right)\,,
\end{equation}
with the constant $\delta_1$ fixed by these boundary conditions to
\begin{equation}
\delta_1=2^{-2/3}b_0^{1/6}\xi_1\,.
\end{equation}

\section{Coefficients of $\dot\rho_s$ in planar $\mathcal{N}=4$ SYM theory}
\label{app:rhos-exp}

In this appendix, we compute the $\alpha_s$ expansion of the coefficients in the development of $\dot\rho_s$ given by Eq.\,\eqref{eq:rhos_nonlinear_fixed}:
\begin{equation}
\dot\rho_s(Y)=c+\frac{\delta_1}{Y}+\frac{\delta_2}{Y^{3/2}}+\mathcal{O}\left(\frac{1}{Y^2}\right)\,,
\end{equation}
in the planar limit of the conformal $\mathcal{N}=4$ SYM theory.
At order $\alpha_s^{3/2}$, we have from the identities Eqs.\,\eqref{eq:cNLL} and \eqref{eq:betacNLL},
\begin{align}
c&= 1+2\sqrt{\abar}+2\abar+\left(1-\frac{\pi^2}{6}\right)\abar^{3/2}+\mathcal{O}(\abar^2)\,,\\
\beta_c&=\sqrt{\abar}-\abar+\left(\frac{1}{2}-\frac{\pi^2}{4}\right)\abar^{3/2}+\mathcal{O}(\abar^2)\,.
\end{align}
Therefore, the coefficient of the $1/Y$ term in $\dot\rho_s$ reads
\begin{align}
\delta_1&=\frac{3}{2(\beta_c-1)}\,,\\
&=-\frac{3}{2}-\frac{3}{2}\sqrt{\abar}+\left(\frac{3}{4}+\frac{3\pi^2}{8}\right)\abar^{3/2}+\mathcal{O}(\abar^2)\,,
\end{align}
and the coefficient of the $1/Y^{3/2}$ term is given by 
\begin{align}
\delta_2&=\frac{3}{2(\beta_c-1)^2}\sqrt{\frac{2\pi}{\chi''(\beta_c)}}\,,\\
&=\frac{3\sqrt{\pi}}{2\abar^{1/4}}\left[\abar^{1/2}+\frac{1}{2}\abar-\left(\frac{21}{8}-\frac{\pi^2}{8}\right)\abar^{3/2}+\mathcal{O}(\abar^2)\right]\,.
\end{align}
\end{document}